# Deep Learning How to Fit an Intravoxel Incoherent Motion Model to Diffusion-Weighted MRI

Sebastiano Barbieri[1,*], Oliver J. Gurney-Champion[2], Remy Klaassen[3], Harriet C. Thoeny[4]

1 Centre for Big Data Research in Health, UNSW, Sydney, Australia

2 Joint Department of Physics at The Institute of Cancer Research and The Royal Marsden NHS Foundation Trust, London, UK

3 Cancer Center Amsterdam, Department of Medical Oncology and LEXOR (Laboratory for Experimental Oncology and Radiobiology), Academic Medical Center, Amsterdam, The Netherlands

4 Department of Radiology, HFR Fribourg-Hôpital Cantonal, Fribourg, Switzerland

* Corresponding author:

**Name:** Sebastiano Barbieri, PhD

**Department:** Centre for Big Data Research in Health

**Institute:** UNSW

**Address:** UNSW Sydney, NSW, 2052 Australia

**E-mail:** s.barbieri@unsw.edu.au

This work was supported in part by grants from Carigest (Geneva, Switzerland), representing an anonymous donor; Nano-Tera (RTD: 20NA21_145919); Maiores Foundation; Kurt and Senta Herrmann Foundation; Propter Homines Foundation; Cancer Research UK Program Grants C33589/A19727 and C7224/A23275; and Swiss National Science Foundation Nr. 32003B_176229/1.

# Abstract

**Purpose:** This prospective clinical study assesses the feasibility of training a deep neural network (DNN) for intravoxel incoherent motion (IVIM) model fitting to diffusion-weighted magnetic resonance imaging (DW-MRI) data and evaluates its performance.

**Methods:** In May 2011, ten male volunteers (age range: 29 to 53 years, mean: 37 years) underwent DW-MRI of the upper abdomen on 1.5T and 3.0T magnetic resonance scanners. Regions of interest in the left and right liver lobe, pancreas, spleen, renal cortex, and renal medulla were delineated independently by two readers. DNNs were trained for IVIM model fitting using these data; results were compared to least-squares and Bayesian approaches to IVIM fitting. Intraclass Correlation Coefficients (ICC) were used to assess consistency of measurements between readers. Intersubject variability was evaluated using Coefficients of Variation (CV). The fitting error was calculated based on simulated data and the average fitting time of each method was recorded.

**Results:** DNNs were trained successfully for IVIM parameter estimation. This approach was associated with high consistency between the two readers (ICCs between 50 and 97%), low intersubject variability of estimated parameter values (CVs between 9.2 and 28.4), and the lowest error when compared with least-squares and Bayesian approaches. Fitting by DNNs was several orders of magnitude quicker than the other methods but the networks may need to be re-trained for different acquisition protocols or imaged anatomical regions.

**Conclusion:** DNNs are recommended for accurate and robust IVIM model fitting to DW-MRI data. Suitable software is available at (1).

# Introduction

In recent years, there has been a renewed interest in the use of artificial neural networks for data classification and regression analysis. Examples of applications in the medical domain include the prognosis of Alzheimer's disease and mild cognitive impairment (2), the classification of digital images of skin lesions with accuracy comparable to human skin-care specialists (3,4), and the prediction of patient longevity based on routinely acquired computerized tomography images (5). Nonetheless, it remains to be seen whether the identification of strong, but theory-free, associations in clinical data can actually translate into improved clinical care (6).

In a noteworthy parallel development, the Intravoxel Incoherent Motion (IVIM) model for the analysis of diffusion-weighted magnetic resonance imaging (DW-MRI) was originally proposed in the eighties (7) but has reached widespread use in clinical research only recently (8-13). Interest in DW-MRI and particularly in IVIM is expected to increase further due to concerns related to the administration of Gadolinium-based contrast agents (14,15). The IVIM model assumes that signal attenuation in DW-MRI occurs because of both diffusion phenomena and bulk movement of water molecules in predefined structures (e.g. capillary perfusion). Mathematically, it expresses the diffusion-weighted signal S(b) acquired with a specific diffusion-weighting (b-value) as the weighted sum of a pure diffusion component and a perfusion dominated "pseudo-diffusion" component:

$$S(b) = S_0[F_p e^{-bD_p} + (1 - F_p)e^{-bD_t}] \qquad (1)$$

where $D_t$ is the pure diffusion coefficient, $F_p$ is the perfusion fraction, and $D_p$ is the pseudo-diffusion coefficient; $S_0$ is the signal acquired without diffusion-sensitizing gradient.

Despite the advancements reported in clinical research, further technical developments are necessary to increase the reproducibility of IVIM-based readings of DW-MRI and establish the application of IVIM in daily clinical routine (16). For example, results may differ significantly depending on which algorithm is used for fitting the IVIM model (17). Among the different fitting algorithms, a Bayesian approach has been shown to be associated with low inter-subject variability and comparatively high precision and accuracy (17,18); however, it is relatively slow (typically takes hours to fit) and, when weakly informative priors are used, may lead to biased estimates of the pseudo-diffusion coefficients $D_p$ (19). Recent work proposes the use of neural networks for IVIM parameter estimation (20) but is limited by the strong assumption of simulated training and test data being identically distributed. Software that performs precise, accurate, and fast IVIM model fitting to DW-MRI data still appears to be lacking. Thus, the aim of the present study is to assess the feasibility of using unsupervised learning to train a deep neural network (DNN) for IVIM model fitting to DW-MRI data and evaluate its performance.

## Methods

This prospective clinical study was approved by the responsible ethics committees and written informed consent was obtained from all accrued subjects. Analyses of the data of all volunteers have been reported previously (16,17,21,22). In particular, the reproducibility of parameters of mono- and bi-exponential models fitted to DW-MRI in upper abdominal organs was discussed earlier (16,21). The optimal b-value threshold separating diffusion and perfusion effects when fitting an IVIM model was analyzed in another previous study (22). Finally, the variability, precision, and accuracy of six different algorithms commonly used for IVIM fitting were also assessed previously (17). The present manuscript introduces and evaluates the use of DNNs for IVIM fitting.

*Study Population: Volunteers*

Ten male volunteers (age range: 29 to 53 years, mean: 37 years) without any known previous disease affecting their upper abdominal organs were accrued and examined in May 2011 at University Hospital Zurich. Subjects were instructed to fast during the six hours preceding imaging and to drink one liter of water during the last two hours before imaging to minimize variability in their hydration level.

*MR Imaging*

Diffusion-weighted images of the upper abdomen were acquired in the axial plane during free breathing on 1.5T and 3.0T Philips Achieva scanners (Philips Healthcare, Best, The Netherlands) with a 32-channel and a 16-channel flexible anteroposterior phased-array coil, respectively. The MR imagers had gradient amplitudes of 33 mT/m and 40 mT/m and slew rates of 122 T/m/sec and 200 T/m/sec, respectively.

A spin-echo echo planar imaging sequence with the following acquisition parameters was used: time of repetition (TR) 5000 msec for both scanners; time of echo (TE) 71 msec and 56 msec for the 1.5T and 3.0T scanner, respectively; eight b-values (b=0, 10, 20, 60, 150, 300, 500, and 1000 sec/mm$^2$); number of excitations 6; field of view 400 mm × 300 mm; matrix size 128 × 128; section thickness of 5 mm (section gap of 1 mm); and receiver bandwidth of 2788 and 3770 Hz per pixel. A sensitivity encoding parallel acceleration factor of 2 in anteroposterior direction and a spectral selection attenuated inversion recovery fat suppression scheme were used. The chosen set of b-values covers the clinically relevant range in which Gaussian diffusion is observed; further, lower b-values were sampled more densely to better estimate the perfusion fraction $F_p$ and the pseudo-diffusion coefficient $D_p$.

The number of sections ranged from 28 to 39 depending on the imaged subject. The interpolated voxel size was 1.56 mm × 1.56 mm × 5.00 mm. The total acquisition time was approximately 12 minutes.

*Image Analysis*

Two readers (a radiologist with 4 years of experience in body MR imaging and a medical student with specific training in MR imaging anatomy) independently placed circular regions of interest (ROIs) in the left and right liver lobe, pancreas, spleen, renal cortex, and renal medulla using the b=0 sec/mm$^2$ images. The ROIs included a total of 10,340 voxels, both for the images acquired at 1.5T and the images acquired at 3.0T. Further details on ROI sizes can be found in (21). Unsupervised learning was used to train DNNs for IVIM model fitting to DW-MRI data. Fitting results obtained by the DNNs were compared to results by a least-squares trust-region algorithm (all parameters constrained to be within 0 and 1) and a Bayesian-probability based approach to IVIM fitting.

The Bayesian approach followed (19) in the use of lognormal priors for the parameters $D_t$ and $D_p$; however, a beta distribution was chosen as a prior for $F_p$ instead of a uniform distribution. Parameters associated with the prior lognormal and beta distributions were determined empirically by fitting these distributions to results by the least-squares algorithm on the considered data. Point estimates of IVIM parameters were obtained by maximum a posteriori probability (MAP). The Bayesian approach occasionally failed to converge; this occurred rarely for voxels within the delineated ROIs or for simulated data (less than 1% of voxels) but relatively frequently (approximately 10% of voxels) for regions with low SNR within a complete image. Whenever the Bayesian approach failed to converge, it returned the corresponding least-squares estimates.

*Deep Neural Network Architecture for IVIM Fitting*

A feed-forward backward-propagation deep neural network was trained to generate estimates of IVIM parameters ($\widehat{D_t}$, $\widehat{F_p}$, and $\widehat{D_p}$). Training is unsupervised and needs to be repeated for datasets with different distributions (e.g. due to different acquisition protocols or imaged anatomic regions). Since the goal is to encode a given dataset, separate training and testing datasets are not required and the network was trained directly on the dataset of interest. A manually determined threshold was applied to the b=0 sec/mm$^2$ image to exclude most background voxels but none of the voxels belonging to the imaged anatomy from training data.

The network is composed of an input layer, three hidden layers, and an output layer. The passthrough input layer is made of neurons which take the normalized diffusion-weighted signal $S(b)/S_0$ sampled at each b-value as input. The three hidden layers are fully connected, with a number of neurons equal to the number of b-values of the data of interest and an exponential linear unit (ELU) activation function (23). The output layer is made of three neurons which hold the estimated parameter values. Initial network weights were set using He initialization (24) or using a previously trained network. The number of hidden layers was determined by grid search, see Supporting Information Figure S1.

An Adam optimizer (25) was used for training with the mean squared error between the observed input $S(b)/S_0$ and the signal $\widehat{S(b)}/S_0$, reconstructed based on Equation (1) and $\widehat{D_t}$, $\widehat{F_p}$, $\widehat{D_p}$, as loss function. Early stopping was implemented by terminating training after the loss function did not improve for ten consecutive iterations. The proposed neural network architecture is essentially an autoencoder (26) with the constraint that the input signal should be encoded by the three IVIM parameters. The network does not impose any restrictions on the range of fitted parameter values.

*Simulations*

The considered algorithms were evaluated further on simulated diffusion-weighted signals. These signals were generated based on Equation (1) with $S_0 = 1500$, b-values of 0, 10, 20, 60, 150, 300, 500, 1000 sec/mm$^2$, and pseudorandom values of $D_t$, $F_p$, and $D_p$. Parameter values were sampled uniformly from the following intervals: diffusion coefficient $D_t$ between 0.5 and 2×10$^{-3}$ mm$^2$/sec, perfusion fraction $F_p$ between 10 and 40%, and pseudo-diffusion coefficient $D_p$ between 0.01 and 0.1 mm$^2$/sec; these intervals cover most parameter values observed in abdominal DW-MRI data (16). To simulate the Rician distribution of magnitude MR data, complex Gaussian noise was added to the diffusion-weighted signals before computing the signal's magnitude (27). The DNN was trained on one million simulated diffusion-weighted signals with noise standard deviation sampled uniformly between 0 and 165. Due to computational time constraints related to the Bayesian approach, the considered algorithms were finally evaluated on 10 batches of 10,000 simulated signals with signal-to-noise ratios (SNR), computed as $S_0$ divided by the noise's standard deviation, ranging between 100 (high SNR) and 10 (low SNR). The performance of multiple DNN models trained on the batches of 10,000 simulated signals at various SNRs was evaluated as well.

*Statistical Analysis*

Since the two readers placed ROIs within areas of relatively homogeneous tissue, a good fitting algorithm should produce smooth parameter maps where variability in ROI placement has only a marginal effect on averaged parameter estimates. To test this, the consistency of measurements between the two readers was assessed by Intraclass Correlation Coefficients (ICC type (3,1)) (28).

Average parameter values were computed for each subject and each anatomical region; intersubject variability of these average parameter values was evaluated using Coefficients of Variation (CV, computed as the sample standard deviation divided by the sample mean). For a direct comparison among the three considered algorithms, CVs were averaged across anatomical regions.

Boxplots of fitting errors on the simulated data were generated and evaluated qualitatively. A fitting algorithm was considered precise if the interquartile error range was small and accurate if the median error was close to zero. In addition, the root-mean-square error (RMSE) of the estimated IVIM parameters was plotted against the SNR of the underlying signals.

All analyses were recomputed for each one of the three considered algorithms (least-squares, Bayesian, DNN). The average fitting time of each algorithm was recorded. The considered algorithms were implemented in Python 3.6.4 and PyTorch 0.4.1. Statistical analyses were carried out in Python and R 3.4.3. All code related to this project is available at (1).

# Results

The deep neural network was trained successfully for IVIM parameter estimation. No negative or extreme parameter values were observed among the DNN fits. Examples of parametric maps computed by the considered algorithms (least-squares, Bayesian, DNN) based on images acquired at 1.5T and 3.0T are presented in Figure 1 and Figure 2, respectively. Parametric maps computed by DNN are more detailed and less noisy than those computed by the least-squares or Bayesian approaches.

*Parameter Differences Across Anatomical Regions*

Boxplots of IVIM parameter values measured in upper abdominal organs (averaged across the two readers) are presented in Figure 3 for images acquired at 1.5T and in Figure 4 for images acquired at 3.0T. Parameters of the empirical priors employed by the Bayesian algorithm are reported in Supporting Information Table 1.

*Measurement Consistency Between Readers*

The DNN approach to IVIM parameter estimation was associated with high consistency between the two readers, especially concerning the pure diffusion coefficient $D_t$ (ICCs of 94% and 97%, see Table 1). The Bayesian approach was associated with the highest consistency for the perfusion fraction $F_p$ (ICCs of 77% and 72%). Results for the pseudo-diffusion coefficient $D_p$ were inconsistent, with ICCs for the DNN approach being highest at 1.5T and lowest at 3.0T. In general, due to the small sample size, confidence intervals for the computed ICCs were large and differences between algorithms were not significant at the 5% level.

*Variability Across Subjects*

In a complementary manner, the DNN approach was also associated with low intersubject variability of estimated parameter values (CVs averaged across anatomical regions are reported in Table 2). Of note, the CVs for $D_p$ were greater than 50 when using the least-squares algorithm, greater than 35 when using the Bayesian algorithm, and only 24.4 at 1.5T and 28.4 at 3.0T when using the DNN approach.

*Fitting Error on Simulated Data*

Boxplots of fitting errors on the simulated diffusion-weighted signals further suggest a comparatively high precision and accuracy of the DNN approach (Figure 5). The relatively constant error of the Bayesian approach at low SNR suggests that, for these signals, estimates are dominated by the prior distribution. The DNN was also associated with the lowest RMSE for all three parameters, with the Bayesian approach performing similarly well (Figure 6). Histograms of all parameter values fitted by the DNN are presented in Supporting Information Figure S2. Training multiple DNN models on batches of 10,000 simulated signals at various SNRs rather than a single model on one million simulated signals led to reduced precision at low SNR and reduced accuracy for $D_t$ estimates, although precision was increased slightly at low SNR (Supporting Information Figure S3); overall, training on a larger dataset is preferable.

*Average Fitting Time*

The average fitting time per voxel for the DNN approach was only $4 \times 10^{-6}$ seconds. The average fitting time was $8 \times 10^{-3}$ seconds for the least-squares algorithm and $9 \times 10^{-2}$ seconds for the Bayesian algorithm. Training time for the DNN was proportional to the amount of training data but was generally under 5 minutes. All computations were carried out on a laptop's CPU (Intel Core i7-6600U CPU at 2.60 GHz).

## Discussion

The present study illustrates the feasibility of training a deep neural network to fit an intravoxel incoherent motion model to diffusion-weighted MRI data. Parametric maps computed by the proposed algorithm were visually improved compared with least-squares and Bayesian approaches. Further, the root-mean-square error of estimated IVIM parameters based on simulated signals with a known ground truth was lowest for the deep neural network. Compared with the Bayesian approach, the proposed algorithm does not require the specification of a prior distribution. Both the neural network and the Bayesian approach were associated with the highest consistency between readers for some of the measured parameter values; however, in the absence of ground truth it is also plausible that the algorithms failed to account for within-organ heterogeneity instead of providing more consistent results.

The use of an artificial neural network to estimate parameters derived from DW-MRI was presented in previous work (20); however, the proposed "shallow" implementation leads to biased estimates of $F_p$ and

$D_p$ as well as varying performance depending on the SNR of the diffusion-weighted signals used for training. The previously applied approach (20) also requires the generation of training data with the same distribution as the considered test data. These problems are obviated in this manuscript by training the network in an unsupervised fashion. Further, early attempts by the authors of this study to train a deep neural network by minimizing errors of estimated parameter values based on simulated training data, similar to (20), led to very narrow spreads in $D_p$ values centered around the mean $D_p$ value in the training data (results not shown); with unsupervised learning this is no longer the case.

Recent work showed that the use of informative (e.g. lognormal) priors increases the precision and accuracy of Bayesian approaches to IVIM fitting, particularly regarding the pseudo-diffusion parameter $D_p$ (19). However, the use of informative priors may find limited acceptance in clinical applications where a high level of objectivity is required. An alternative approach is to deduce the Bayesian prior from a neighborhood of the pixel of interest; nonetheless, assuming a common distribution across pixels may lead to disappearing structures (29). The present study suggests that DNNs may be used to fit IVIM parameters with high precision and accuracy as well as high objectivity. Interestingly, the parameter maps generated by the DNN are very smooth in homogeneous tissues despite being computed independently for each voxel. It is plausible that the DNN learns the manifold of realistic IVIM parameter values and maps observed input signals onto this manifold, thereby reducing noisy parameter estimates (30).

Using DNNs for IVIM fitting is several orders of magnitude quicker than using least-squares or Bayesian methods. Nonetheless, training of the network may need to be repeated for different acquisition protocols or imaged anatomic regions. In clinical software, each imaging protocol could be associated with a specific DNN for IVIM fitting which is re-calibrated at regular intervals outside imaging hours. The consistent integration of IVIM imaging within clinical processes could lead to reduced contrast medium administration and corresponding cost savings. The proposed algorithm could also be implemented directly on MR scanners and lead to automated quality control checks of estimated parameter maps while the patient is still in the MR scanner.

Limitations of the present study include the small sample size of only ten volunteers, which we plan to address in a future prospective study. Additional examples of parametric maps of pancreatic cancer patients are presented in Supporting Information Figure S4. Despite our best efforts, we were not able to improve convergence properties of the Bayesian approach in image regions with low SNR. This may be addressed by using sampling techniques instead of MAP; however, this further increases computational cost and leads to difficulties in assessing convergence. In addition, DNNs were trained again for each dataset of interest and we did not assess whether it is possible to perform training on a set of patients and testing on a different set. However, given the relatively small number of parameters in the network, training takes only a few minutes.

In conclusion, the present study introduces a non-supervised DNN approach to estimate IVIM parameters. Its performance was shown to be comparable to the current state-of-the-art approach (Bayesian) with the advantage of being considerably faster and producing visually improved parametric maps. A Jupyter Notebook with a brief demo of the software to train the neural network and fit the IVIM model to DW-MRI data is available for download at (1). Fellow researchers and clinicians are encouraged to test the software and report their experience.

## Acknowledgements

We would like to thank Olivio F. Donati and Erik Andres for their contribution to image analysis. We would also like to thank our volunteers for participating in the present study. Further, we would like to thank Aart J Nederveen for facilitating the patient scans and Hanneke W.M. van Laarhoven and Johanna W. Wilmink for sharing patient data.

# Figures and Tables

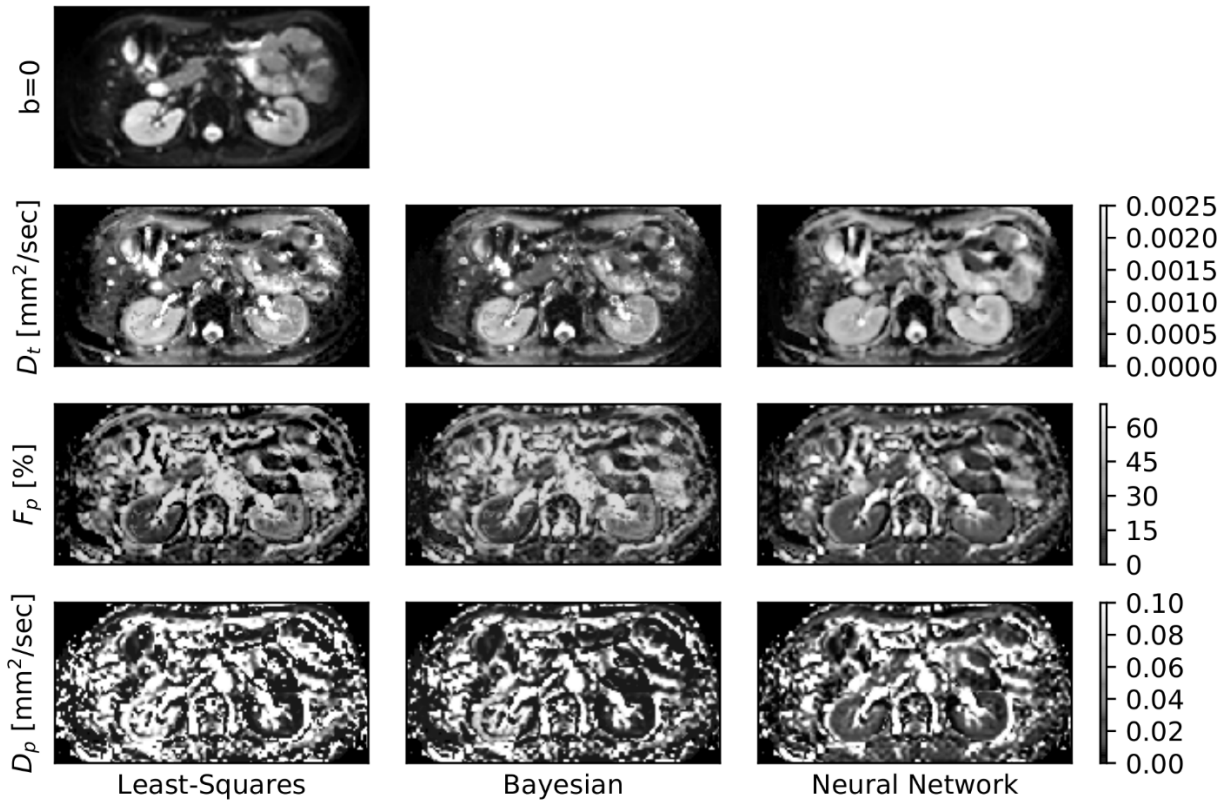

**Figure 1.** Axial $S_0$ sample image of the upper abdomen of a 38-year-old male volunteer at the midlevel of the kidneys with corresponding parametric maps of the pure diffusion coefficient $D_t$, the perfusion fraction $F_p$, and the pseudo-diffusion coefficient $D_p$. Imaging was performed at 1.5T. Note that, for improved visualization in this figure, parameter values fitted by least-squares and Bayesian approaches were restricted to the following intervals: 0 to 0.005 mm$^2$/sec for $D_t$, 0 to 60 % for $F_p$, and 0.01 to 0.30 mm$^2$/sec for $D_p$. In the parametric maps computed by the neural network the outer contours of the kidneys are delineated better, and the renal parenchyma is more homogeneous. Further, parameter values computed by DNN are similar in the right and the left kidney, as expected in a healthy volunteer.

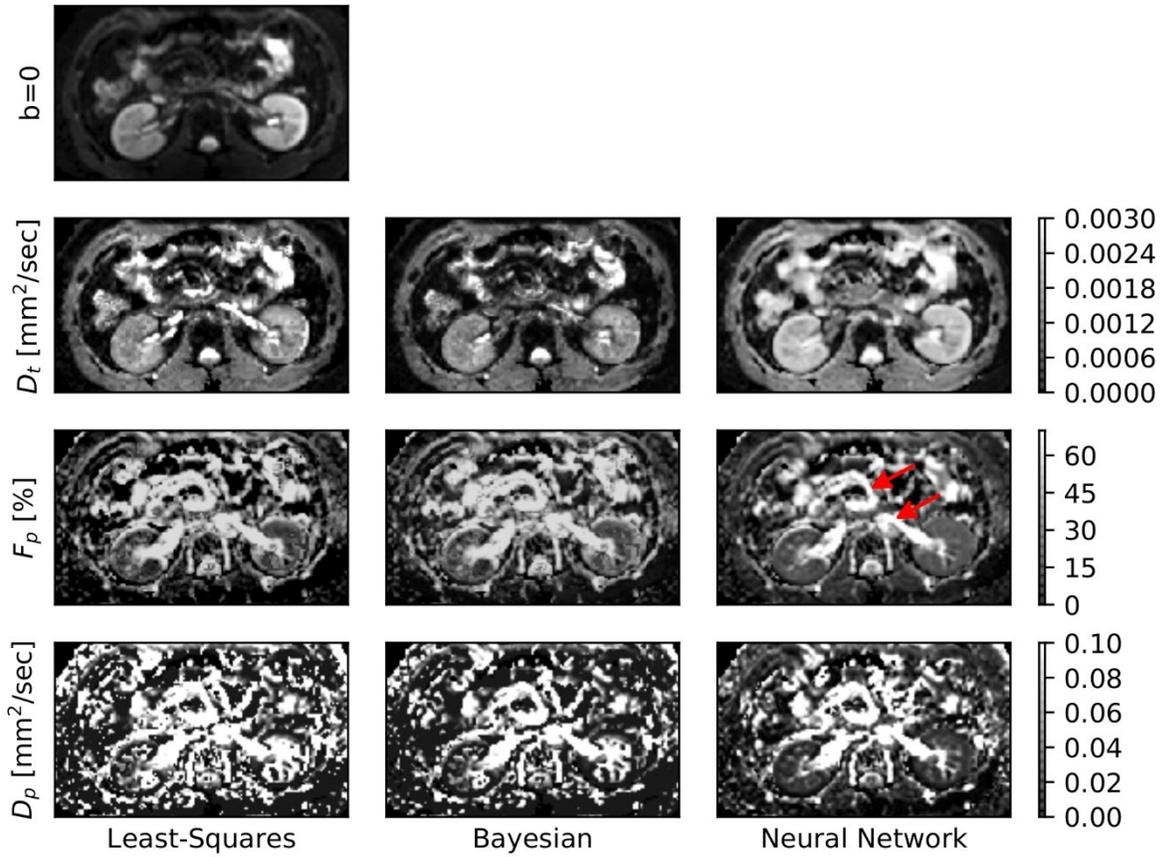

**Figure 2.** Axial $S_0$ sample image of the upper abdomen of a 30-year-old male volunteer at the midlevel of the kidneys with corresponding parametric maps of the pure diffusion coefficient $D_t$, the perfusion fraction $F_p$, and the pseudo-diffusion coefficient $D_p$. Imaging was performed at 3.0T. Note that, for improved visualization in this figure, parameter values fitted by least-squares and Bayesian approaches were restricted to the following intervals: 0 to 0.005 mm$^2$/sec for $D_t$, 0 to 60 % for $F_p$, and 0.01 to 0.30 mm$^2$/sec for $D_p$. Similarly to Figure 1, in the parametric maps computed by the neural network the outer contours of the kidneys are delineated better, the renal parenchyma is more homogeneous, and parameter values are similar in the right and the left kidney. In addition, the $D_t$ map computed by the neural network facilitates the differentiation between renal cortex and renal medulla and abdominal vessels are delineated better on the $F_p$ map (arrows).

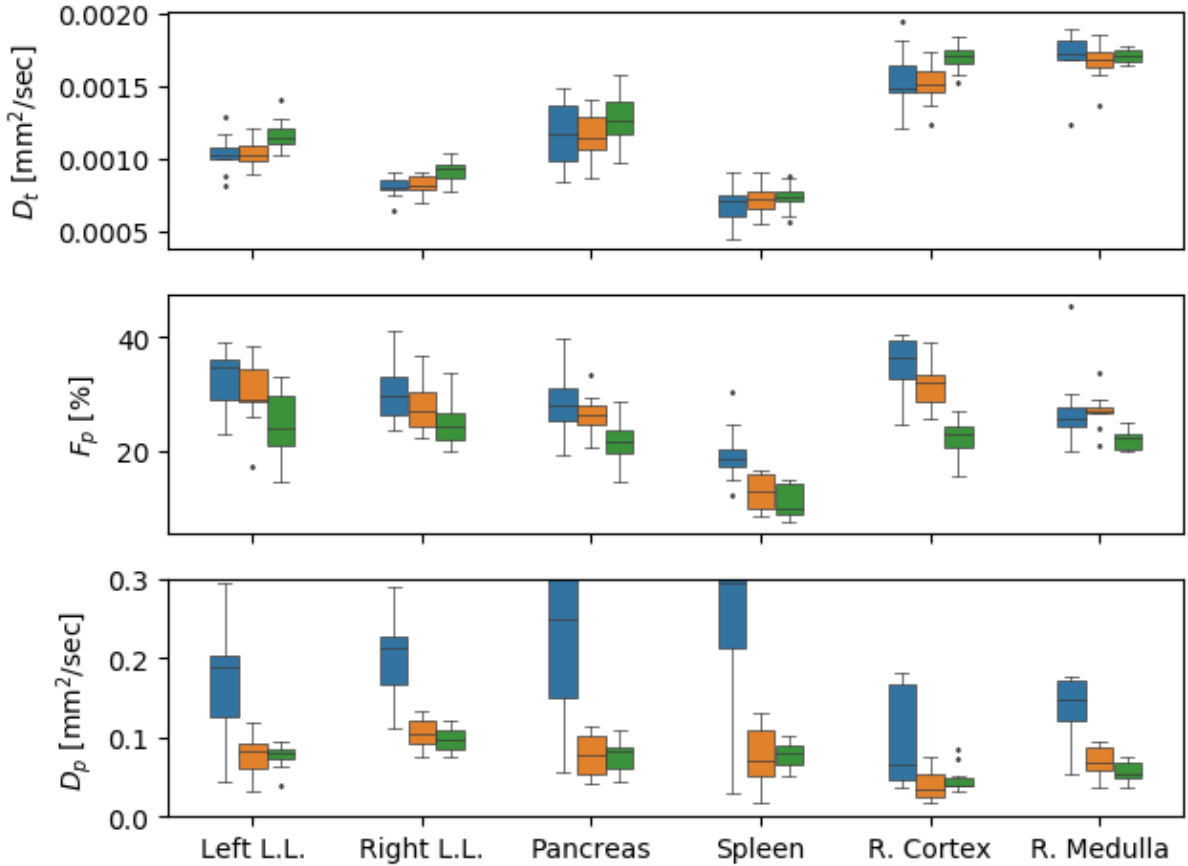

**Figure 3.** Boxplots of intravoxel incoherent motion parameters ($D_t$, $F_p$, and $D_p$) fitted in upper abdominal organs (LL, liver lobe; R, renal) by the considered algorithms (■=Least-Squares, ■=Bayesian, ■=DNN) based on images acquired at 1.5T. The central marks are the medians and the boxes extend from the first ($Q_1$) to the third ($Q_3$) data quartile; data points further away than 1.5 times the distance $Q_3$-$Q_1$ are considered outliers.

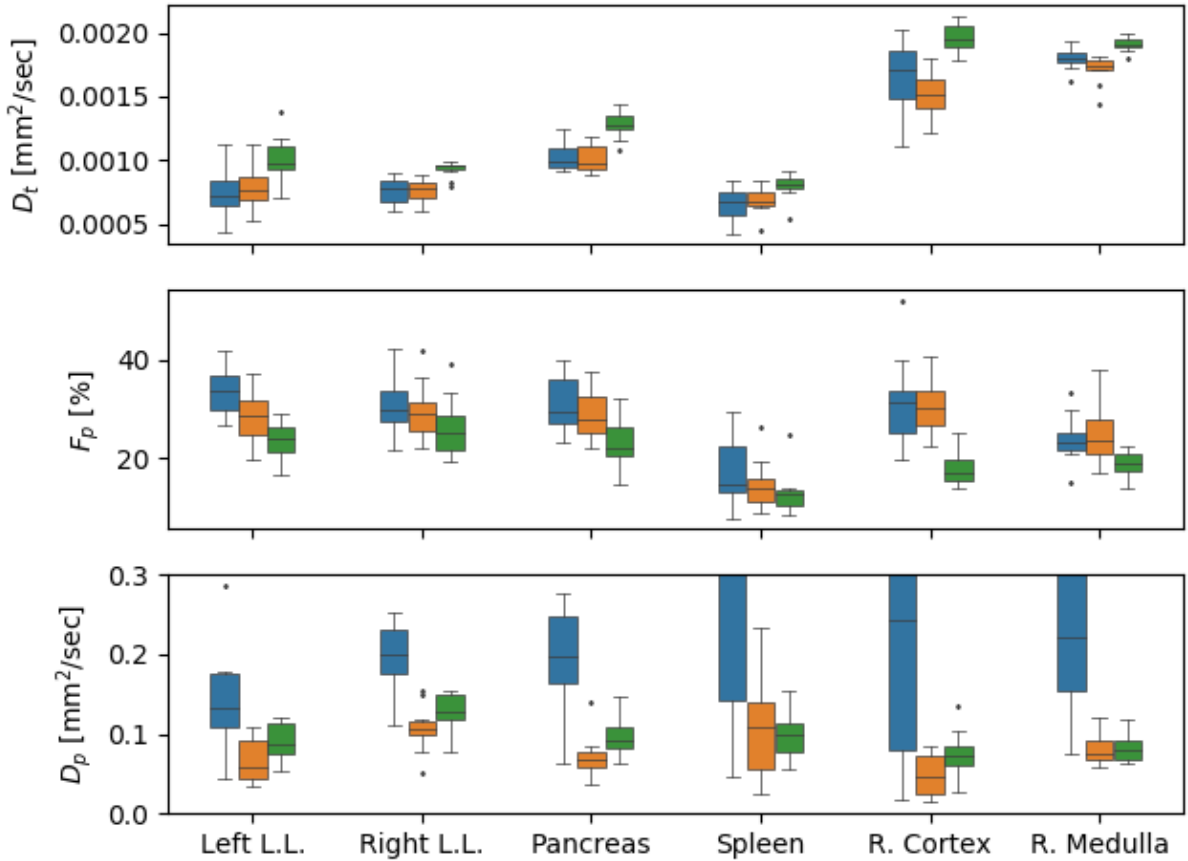

**Figure 4.** Boxplots of intravoxel incoherent motion parameters ($D_t$, $F_p$, and $D_p$) fitted in upper abdominal organs (LL, liver lobe; R, renal) by the considered algorithms (■=Least-Squares, ■=Bayesian, ■=DNN) based on images acquired at 3.0T. The central marks are the medians and the boxes extend from the first ($Q_1$) to the third ($Q_3$) data quartile; data points further away than 1.5 times the distance $Q_3$-$Q_1$ are considered outliers.

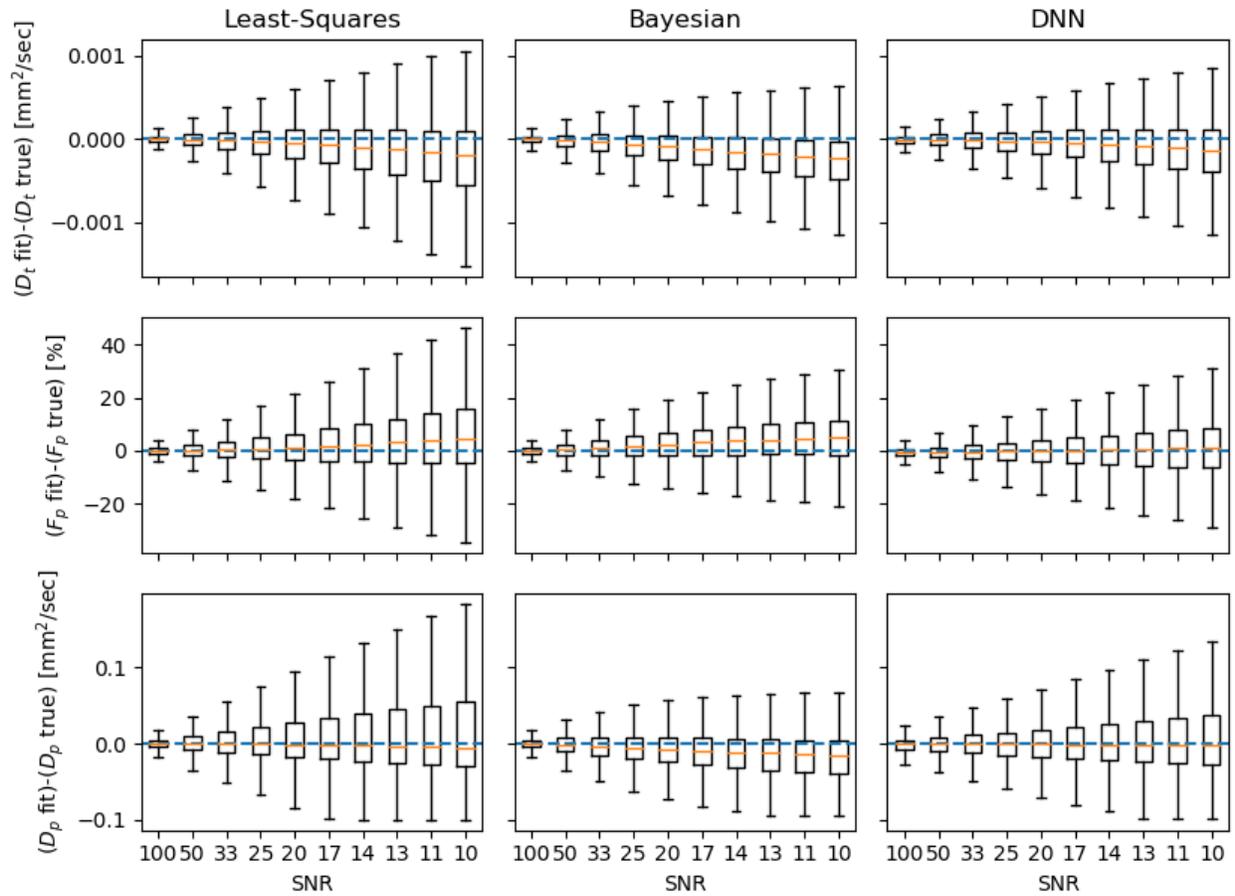

**Figure 5.** Boxplots of errors associated with the considered algorithms (Least-Squares, Bayesian, DNN) when fitting an intravoxel incoherent motion model to 10,000 simulated diffusion-weighted signals at signal-to-noise ratios (SNR) of 100 (boxes on the left), 50, 33, …, 11, 10 (boxes on the right). The central marks are the medians and the boxes extend from the first ($Q_1$) to the third ($Q_3$) data quartile. For visual clarity, outliers are not shown. The blue dashed lines correspond to an error of zero.

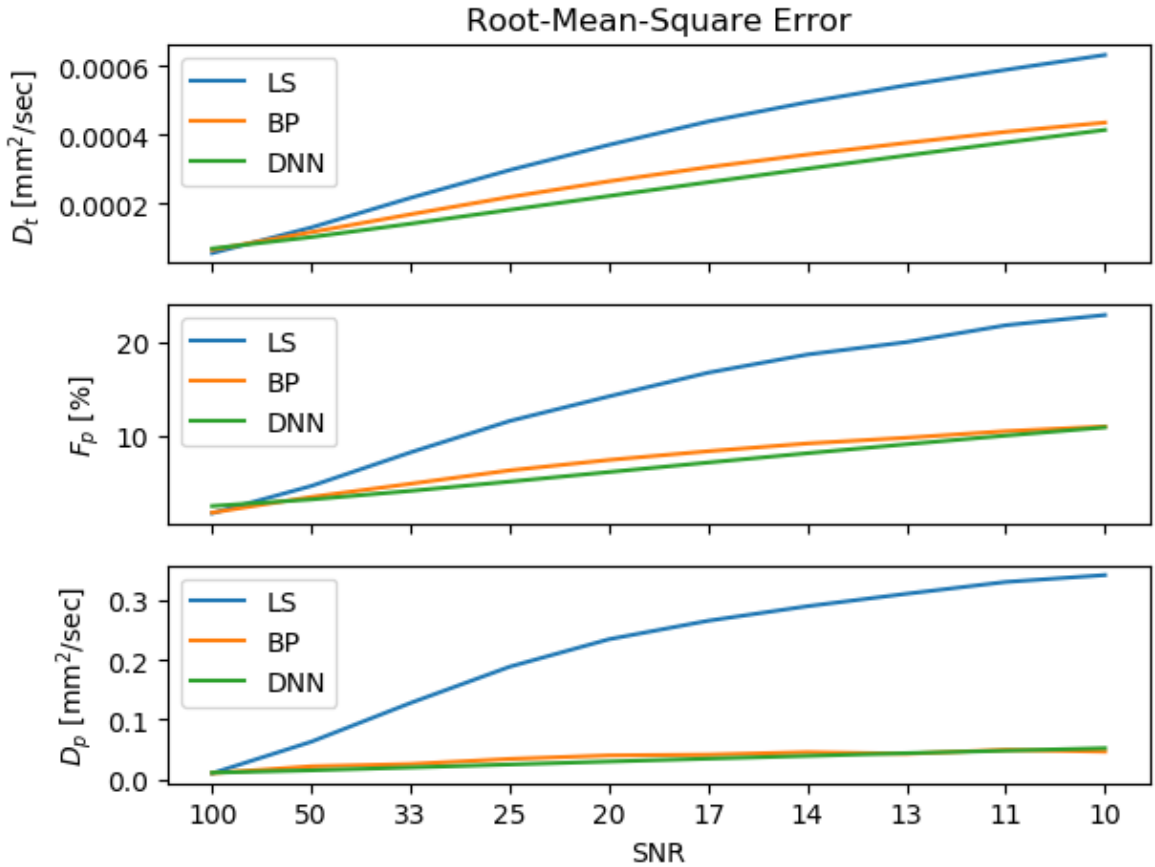

**Figure 6.** Plots of root-mean-square error (RMSE) of the estimated intravoxel incoherent motion parameters ($D_t$, $F_p$, and $D_p$) vs signal-to-noise ratio (SNR) for the three considered algorithms (LS: Least-Squares, BP: Bayesian, DNN: Deep Neural Network).

|  | ICC(3,1) [%] [95% Confidence Interval] | | | | | |
|---|---|---|---|---|---|---|
|  | 1.5T | | | 3.0T | | |
| Parameter | Least-Squares | Bayesian | DNN | Least-Squares | Bayesian | DNN |
| $D_t$ | 88 [81, 93] | 92 [87, 95] | 94 [89, 96] | 92 [87, 95] | 96 [93, 97] | 97 [96, 98] |
| $F_p$ | 63 [45, 76] | 77 [65, 86] | 66 [50, 79] | 57 [37, 72] | 72 [57, 82] | 66 [49, 78] |
| $D_p$ | 36 [12, 56] | 38 [15, 58] | 50 [28, 66] | 71 [56, 82] | 70 [54, 81] | 51 [29, 67] |

**Table 1.** Intraclass Correlation Coefficients (in percentage) with corresponding 95% confidence intervals for measurements by two readers in upper abdominal organs (left and right liver lobe, pancreas, spleen, renal cortex, and renal medulla). Analyses were repeated for images acquired at 1.5T and 3.0T and for each one of the considered algorithms (Least-Squares, Bayesian, DNN: Deep Neural Network).

|  | CV [%] | | | | | |
|---|---|---|---|---|---|---|
|  | 1.5T | | | 3.0T | | |
| Parameter | Least-Squares | Bayesian | DNN | Least-Squares | Bayesian | DNN |
| $D_t$ | 14.5 | 10.9 | 9.3 | 16.0 | 13.2 | 9.2 |
| $F_p$ | 20.2 | 17.1 | 18.6 | 24.2 | 22.8 | 22.3 |
| $D_p$ | 52.7 | 35.5 | 24.4 | 53.0 | 41.5 | 28.4 |

**Table 2.** Intersubject Coefficients of Variation of $D_t$, $F_p$, and $D_p$, averaged across anatomical regions (left and right liver lobe, pancreas, spleen, renal cortex, and renal medulla). Analyses were repeated for images acquired at 1.5T and 3.0T and for each one of the considered algorithms (Least-Squares, Bayesian, DNN: Deep Neural Network).

# Supporting Information

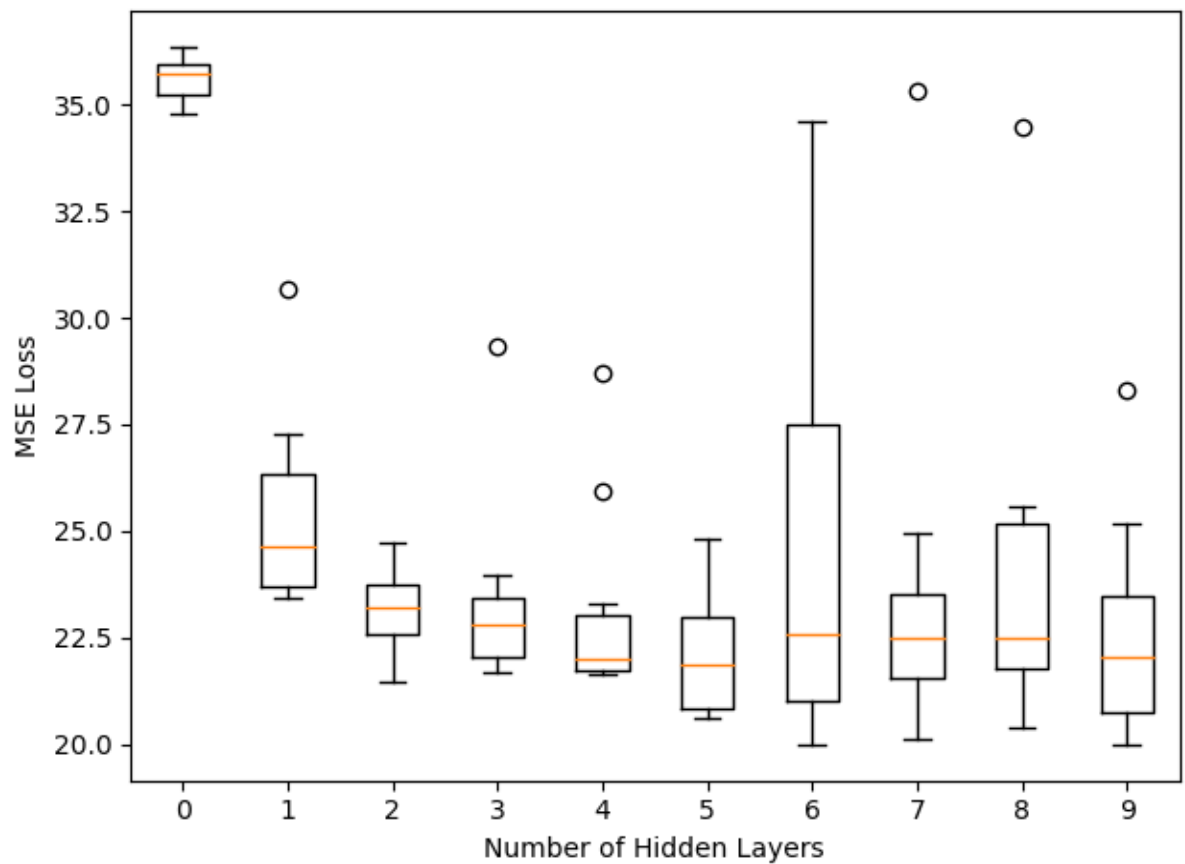

**Supporting Information Figure S1.** Boxplots of cumulated mean squared error (MSE) loss of the neural network trained on one million simulated diffusion-weighted signals vs. the number of hidden layers of the network. For each number of hidden layers, training of the network was repeated 10 times with different random parameter initializations and order of mini-batches. Adding more than two or three hidden layers to the neural network has a limited effect on reducing MSE loss and may introduce unnecessary complexity in the network.

| Parameter | 1.5T | 3.0T |
|---|---|---|
| Lognormal prior of $D_t$, *shape* | 0.554 | 0.558 |
| Lognormal prior of $D_t$, *scale* | 0.000993 | 0.000909 |
| Beta prior of $F_p$, *a* | 1.04 | 1.26 |
| Beta prior of $F_p$, *b* | 2.66 | 3.31 |
| Lognormal prior of $D_p$, *shape* | 1.54 | 1.49 |
| Lognormal prior of $D_p$, *scale* | 0.0460 | 0.0505 |

**Supporting Information Table 1.** Parameters of the empirical priors employed by the Bayesian algorithm. Further information on these parameters is reported in the SciPy documentation (1,2).

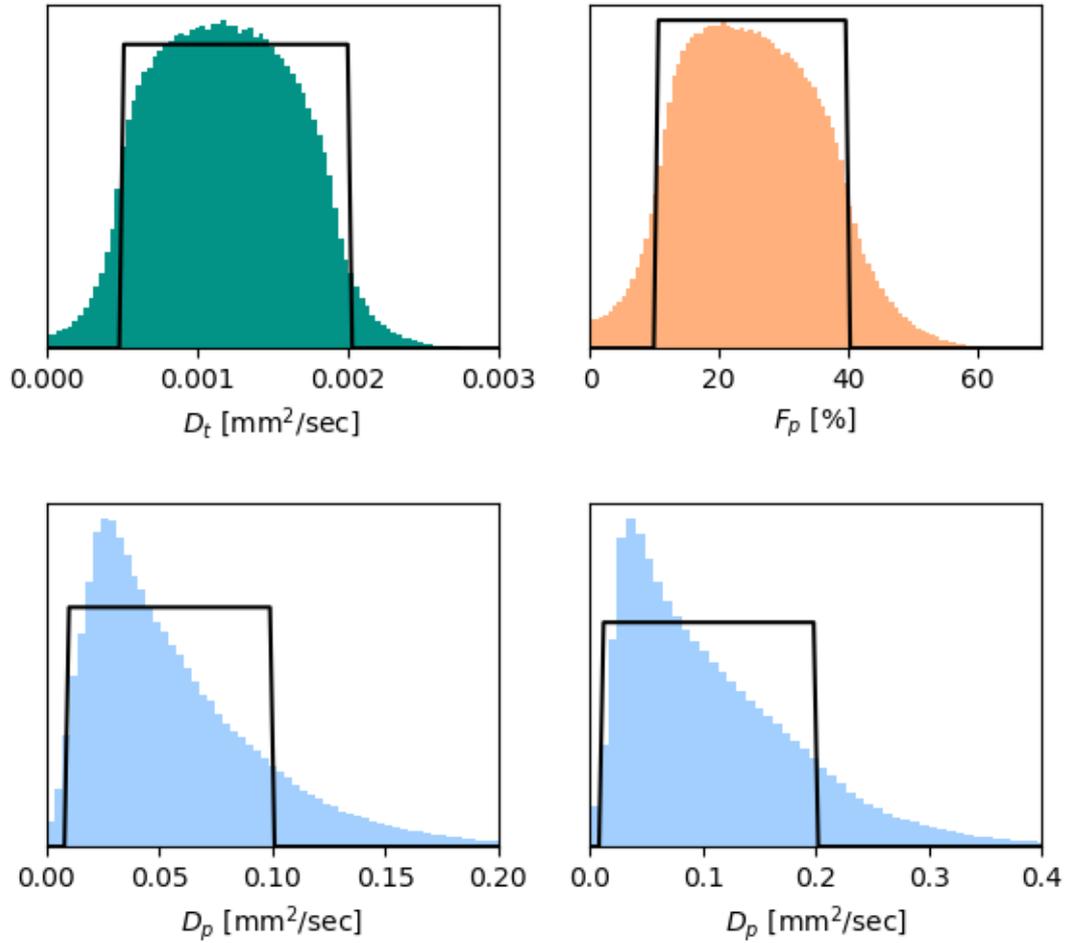

**Supporting Information Figure S2.** Histograms of intravoxel incoherent motion parameters ($D_t$, $F_p$, and $D_p$) fitted by the deep neural network based on simulated diffusion-weighted signals at different signal to noise ratios compared with their true underlying uniform distribution: $D_t$ was sampled from $[0.5, 2] \times 10^{-3}$ mm$^2$/sec, $F_p$ from $[10, 40]$%, and $D_p$ from $[0.01, 0.1]$ mm$^2$/sec (lower left) or from $[0.01, 0.2]$ mm$^2$/sec (lower right).

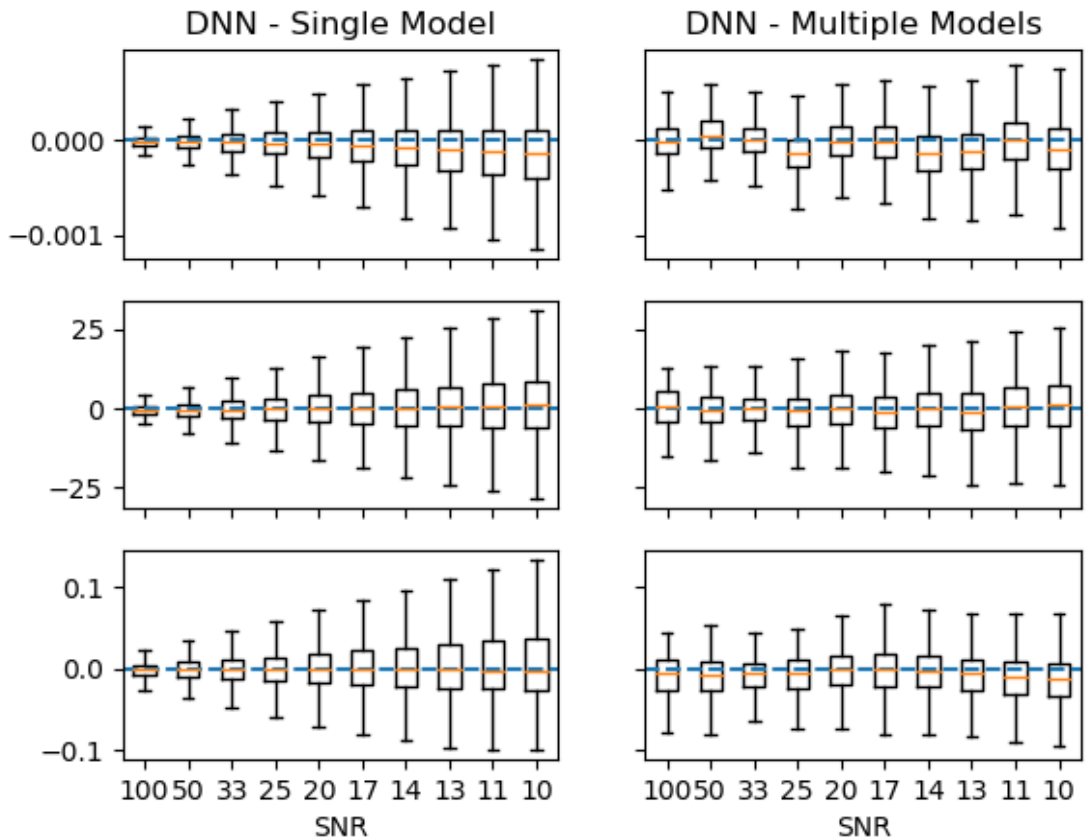

**Supporting Information Figure S3.** Boxplots of errors associated with single and multiple model DNNs when fitting an intravoxel incoherent motion model to 10,000 simulated diffusion-weighted signals at signal-to-noise ratios (SNR) of 100 (boxes on the left), 50, 33, …, 11, 10 (boxes on the right). The single model DNN was trained on one million simulated signals with SNRs between infinity and 9. Each of the multiple DNN models was trained on one of the batches of 10,000 simulated signals with specific SNRs. The central marks are the medians and the boxes extend from the first ($Q_1$) to the third ($Q_3$) data quartile. For visual clarity, outliers are not shown. The blue dashed lines correspond to an error of zero. Parameter estimates computed by the multiple DNN models were less precise at low SNR and occasionally less accurate (particularly for $D_t$); however, they were slightly more precise at low SNR.

## Clinical Application: Imaging of Pancreatic Cancer

As an example of how the trained DNNs might be used in clinical practice, sample parametric maps of five patients with metastatic pancreatic ductal adenocarcinoma (confirmed by histopathology), are presented in Supporting Information Figure S4. Patients were scanned between July 2015 and August 2017 at Amsterdam University Medical Center, location Academic Medical Center, as part of a prospective study (NCT02358161). Diffusion-weighted images were acquired on a 3-T MR Philips Ingenia scanner with b-values of 0, 10, 20, 30, 40, 50, 75, 100, 150, 250, 400, and 600 sec/mm$^2$ (3). Detailed imaging parameters are reported in (4,5). Parametric maps computed by DNN were less noisy, allowing the visualization of additional features within pancreatic lesions and facilitating the localization of metastases in the liver, particularly in the $D_p$ maps.

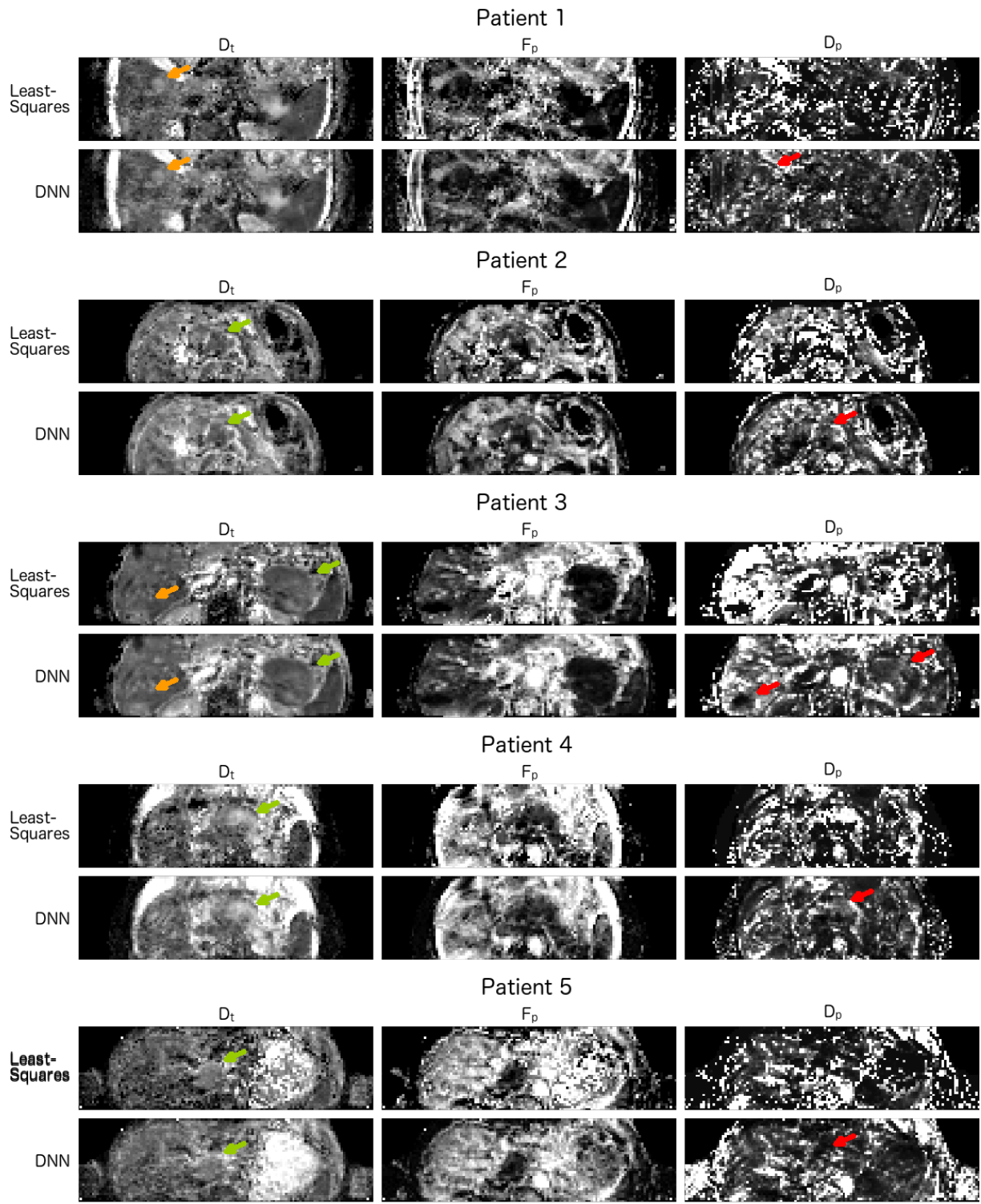

**Supporting Information Figure S4.** Exemplary axial parametric maps of patients with metastatic pancreatic ductal adenocarcinoma, computed by least-squares and by the deep neural network. Pancreatic lesions (green arrows) and metastases in the liver (orange arrows) are visible as hypointense regions on maps of the pure diffusion coefficient $D_t$. Red arrows point to cancer on maps of the pseudo-diffusion coefficient $D_p$; when least-squares fitting is used, visibility is hampered by image noise and nearby hyperintense regions. Note that, for improved visualization in this figure, parameter values fitted by least-squares were restricted to the following intervals: 0 to 0.005 mm$^2$/sec for $D_t$, 0 to 60 % for $F_p$, and 0.01 to 0.30 mm$^2$/sec for $D_p$. DW-MRI data was acquired at 3.0T. *Patient 1* is a 50-year-old male with pancreatic cancer, the figure shows an untreated liver metastasis. *Patient 2* is a 64-year-old female with multiple lymph node and liver metastasis after whipple resection. *Patient 3* is a 74-year-old male with a locally advanced pancreatic tail tumor. Note the improved visibility of the metastasis on the $D_t$ map computed by the DNN. *Patient 4* is a 77-year-old female with a tumor in the pancreatic tail. Note the enhanced image quality of $D_p$ from the DNN compared to the least-squares fits. *Patient 5* is a 48-year-old male with liver metastasis after whipple resection.